\documentclass[fleqn,twoside]{article}
\usepackage{espcrc2}
\usepackage{graphicx}

\title{Daya Bay Neutrino Experiment\thanks{http://dayawane.ihep.ac.cn/}}

\author{Jun Cao\address {Institute of High Energy Physics,
         P.O. Box 918, Beijing 100049, China}
        }

\begin{document}

\begin{abstract}
The Daya Bay Neutrino Experiment is proposed to measure
$\sin^22\theta_{13}$ to better than 0.01 at 90\% C.L. in a
three-year run. The experimental site, detector design, and
background estimation are presented.
\end{abstract}

\maketitle

\section{Experimental Goal}
The Daya Bay Neutrino Experiment is a neutrino oscillation
experiment designed to measure the mixing angle $\theta_{13}$
using anti-neutrinos produced by the reactors of the Daya Bay
Nuclear Power Plant (NPP) and the LingAo NPP. The experiment is
proposed to measure $\sin^22\theta_{13}$ to better than 0.01 at
90\% confidence level in a three-year run, from 2009 to 2011.

\section{The Site}
Daya Bay NPP and LingAo NPP locate in the south of China, 55 km to
the northeast of Hong Kong and 45 km to the east of Shen Zhen
city. The two NPPs are about 1100 m apart. Each NPP has two cores
running. Another two cores, called LingAo II, are expected to
commission in 2010. The thermal power of each core is 2.9 GW.
Hence the existing total thermal power is 11.6 GW, and will be
17.4 GW after 2010.

\begin{figure}[ht]
\begin{center}
\includegraphics[width=0.45\textwidth]{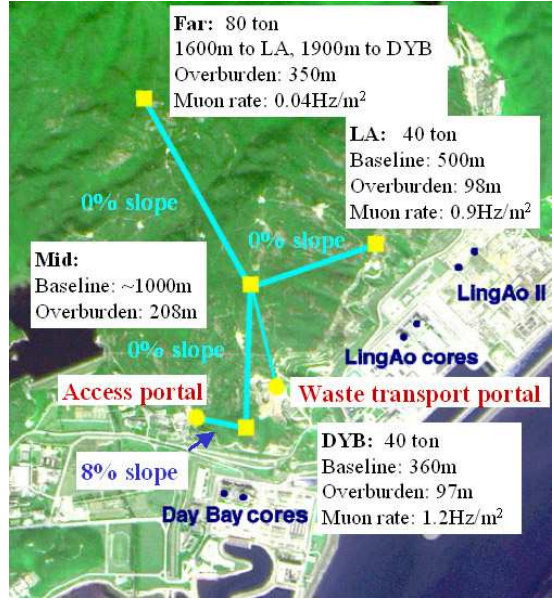}
\caption{Layout of the Daya Bay experiment.}\label{fig:layout}
\vspace{-0.1in}
\end{center}
\end{figure}

The NPPs are adjacent to high hills that can provide protection
from cosmic rays that are the main source of backgrounds in the
experiment. A horizontal tunnel will be built to connect the
underground experimental halls. The tunnel layout approved by the
NPPs is shown in Fig.~\ref{fig:layout}.

For the near-far relative measurement, the reactor errors can be
reduced by more than an order, comparing to a single-detector
measurement. If the uncorrelated error of a single core is 2\%,
the residual reactor error is $\sim$ 0.06\%.

Daya Bay is a versatile site. A fast measurement could be
conducted with a single DYB near site and a mid site during tunnel
construction. The sensitivity of $\sin^22\theta_{13}$ could reach
$\sim$ 0.03 in an one-year run. A mid-far measurement could be
used to cross check the near-far measurement with different
systematic.

\section{Detector}
Each detector site has a water buffer to provide greater than 2
meters of shielding to reduce cosmogenic neutrons from the rock.
The water buffer serves also as a Cherenkov detector to veto
cosmic muons. Outside the water buffer, there is another muon veto
detector, either plastic scintillator or Resistive Plate Chamber
(RPC). The combined veto efficiency is expected to be $>$ 99.5\%.
Identical main detector modules are placed inside the water
buffer, two at each near site and four at the far site. The module
consists of three layers. The inner layer is filled with 20 ton
Gd-doped liquid scintillator, surrounded by $\sim$ 45 cm normal
liquid scintillator in the middle layer. The outer layer is filled
with $\sim$ 45 cm mineral oil. PMTs are mounted in the oil.

The detector is simulated with a GEANT3 based Monte Carlo program,
followed by event reconstruction. The neutron selection efficiency
with a 6 MeV cut is $\sim$ 92\%. The efficiency error caused by
1\% energy scale error is 
 $\sim$ 0.2\%. The
inefficiency of the positron selection with a 1 MeV cut can be
ignored, as well as its error.

The detector module is designed to be movable. The filled module,
together with its electronics, will be moved as a whole during the
detector swapping.

\section{Backgrounds}
The accidental background, the fast neutron background, and the
$^8$He/$^9$Li isotopes are three major backgrounds in the
experiment. Cosmic muons are simulated using the MUSIC package,
with the detailed contour map of the mountain.

The isotope backgrounds are calculated with the cross section
measured at CERN \cite{hagner}. They are dominated by $^9$Li
\cite{kamland} and can be measured in-situ by fitting the time
since last muon with a precision approximated to be $
\sigma=\frac{1}{\sqrt{N}} \sqrt{(1+\tau R_\mu)^2-1}$, where $N$ is
the number of neutrino candidates, $\tau$ is the lifetime of
$^9$Li, and $R_\mu$ is the muon rate in the detector.

The fast neutron background and the rate of neutron singles are
estimated by a full Monte Carlo simulation with the neutron
production calculated by an empirical formula \cite{wang}.
The natural radioactivities are also estimated by a full Monte
Carlo simulation, with the PMT glass, surrounding rock, radon in
the air, etc. taken into account. The accidental and fast neutron
backgrounds can be inferred from in-situ measurements. The
backgrounds present in the Daya Bay experiment are summarized
below.
\begin{center}
\begin{tabular}{ccc}
 \hline\hline
  & Near Site & Far Site \\
 Accidentals B/S    & $<$0.05\% & $<$0.05\% \\
 Fast Neutron B/S   & 0.15\%    & 0.1\%   \\
 $^8$He/$^9$Li B/S  & 0.55\%    & 0.25\% \\\hline
\end{tabular}
\end{center}

\section{Systematics and Sensitivity}
Detector swapping between the near and far sites will cancel most
detector systematic errors. The residual error is estimated to be
$\sim$ 0.2\%, arising mainly from the energy scale uncertainties.
The detector systematic error and the variation before and after
swapping could be monitored to a certain precision by the
side-by-side calibration of detectors at the near sites.

\begin{figure}[ht]
\begin{center}
\includegraphics[width=0.4\textwidth]{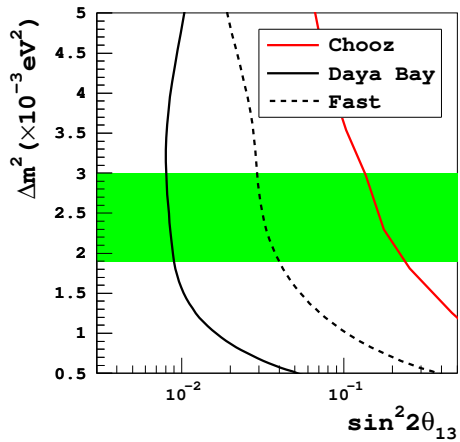}
\caption{Sensitivity of the fast and full measurements at the 90\%
confidence level.}\label{fig:sens} \vspace{-0.2in}
\end{center}
\end{figure}

The expected sensitivity for the fast measurement and the full
operation is shown in Fig.~\ref{fig:sens}.

\end{document}